\input harvmac.tex
%\input epsf.tex
%\draft
%\newcount\figno
\figno=0
\def\fig#1#2#3{
\par\begingroup\parindent=0pt\leftskip=1cm\rightskip=1cm\parindent=0pt
\baselineskip=11pt
\global\advance\figno by 1
\midinsert
\epsfxsize=#3
\centerline{\epsfbox{#2}}
\vskip 12pt
{\bf Fig. \the\figno:} #1\par
\endinsert\endgroup\par
}
\def\figlabel#1{\xdef#1{\the\figno}}
\def\encadremath#1{\vbox{\hrule\hbox{\vrule\kern8pt\vbox{\kern8pt
\hbox{$\displaystyle #1$}\kern8pt}
\kern8pt\vrule}\hrule}}

\overfullrule=0pt

%macros
%

\def\np#1#2#3{{\it Nucl. Phys.} {\bf B#1} (#2) #3}
\def\pl#1#2#3{{\it Phys. Lett. }{\bf B#1} (#2) #3}

\def\physrev#1#2#3{{\it Phys. Rev.} {\bf D#1} (#2) #3}

\font\zfont = cmss10 %scaled \magstep1

\def\bigone{\hbox{1\kern -.23em {\rm l}}}
\def\ZZ{\hbox{\zfont Z\kern-.4emZ}}

\def\a{\alpha}

\def\g{\gamma}
\def\d{\delta}
\def\e{\epsilon}

\def\k{\kappa}
\def\l{\lambda}
\def\m{\mu}
\def\n{\nu}

\def\s{\sigma}
\def\t{\tau}

\def\L{\Lambda}

\def\o{\over}

\def\bbox{{\sqcap \!\!\!\! \sqcup}}

\Title{CALT-68-2149, hep-th/9712238,}
{\vbox{
\hbox{\centerline{On Graviton Scattering Amplitudes in M-Theory}}
\hbox{\centerline{ }}
}}
\smallskip
\centerline{Katrin Becker\footnote{$^\diamondsuit$}
{\tt beckerk@theory.caltech.edu} and Melanie 
Becker\footnote{$^\star$}
{\tt mbecker@theory.caltech.edu} }
\smallskip
\centerline{\it California Institute of Technology 452-48, 
Pasadena, CA 91125}
\bigskip
\baselineskip 18pt
\noindent

We compute graviton scattering amplitudes in M-theory
using Feynman rules for a scalar particle coupled to gravity in 
eleven dimensions.
The processes that we consider describe the single graviton
exchange and the double graviton exchange, that in
M(atrix) theory correspond to the $v^4/r^7$-
and $v^6/r^{14}$-term respectively.
We further show 
that the $ v^6/r^{14}$-term
appearing in M(atrix) theory at two-loops can be obtained from 
the covariant eleven-dimensional
four-graviton amplitude.
Finally, we calculate the $v^8/r^{18}$-term appearing at two-loops in 
M(atrix) theory. It has been previously conjectured that this term is
related to   
a four graviton scattering amplitude
involving the $R^4$-vertex of M-theory.

\Date{December, 1997}
\newsec{Introduction}
M-theory is the `quantum version' of eleven-dimensional supergravity.
The low energy degrees of freedom and interactions of this theory are
those of general relativity.
However, if one wants to calculate e.g. radiative corrections to graviton
scattering amplitudes one has to know how to describe the short distance
degrees of freedom of M-theory.
M(atrix) theory 
\ref\bfss{T.~Banks, W.~Fischler, S.~H.~Shenker and 
L.~Susskind, ``M Theory as a Matrix Model: A Conjecture'', 
\physrev {55} {1997} {5112}, hep-th/9610043. }
is a proposal that in the infinite momentum frame 
these degrees of freedom can be described 
in terms of Dirichlet zero branes. 
Graviton scattering amplitudes that are usually hard to compute
(or even infinite) using standard quantum field theory techniques
can be calculated in terms of a simple quantum mechanical model.
So for example, the $v^4/r^7$-term appearing at one-loop in M(atrix) theory
has an interpretation as a single graviton exchange diagram in eleven
dimensions, while the $v^6/r^{14}$-term appearing at two-loops in 
M(atrix) theory
describes a correction coming from general relativity
\ref \bbpt{K.~Becker, M.~Becker, J.~Polchinski and A.~Tseytlin,
''Higher Order Graviton Scattering in M(atrix) Theory'', \physrev {56} {1997}
{3174}, hep-th/9706072. }.
An example of a quantum gravity correction is the 
$v^8/r^{18}$-term appearing at two-loops in 
M(atrix) theory
\ref\leta{L.~Susskind, Talk at Strings `97.}.

In this note we would like to consider these three terms
in the effective potential between two gravitons in some detail.
In section 2 we will compute the $v^4/r^7$- and $v^6/r^{14}$-terms using 
Feynman rules for supergravity.
In section 3 we will show that the relativistic correction can be calculated
using the covariant four-graviton amplitude in eleven dimensions
\ref\gsb{M.~B.~Green, J.~H.~Schwarz and L.~Brink, ``$N=4$ Yang-Mills
and $N=8$ Supergravity as Limits of String Theories'', 
\np {198} {1982} {474}.} 
\ref\tsru{J.~G.~Russo and A.~A.~Tseytlin, ``One Loop Four Graviton
Amplitude in Eleven-Dimensional Supergravity'', Imperial College preprint, 
IMPERIAL-TP-96-97-53, hep-th/9707134.}.
In section 4 we will make some remarks about quantum gravity corrections.
The appendix contains more explicitly some of our calculations.

\newsec{Single Graviton Exchange}
In this section we would like to consider the scattering of two gravitons
in eleven dimensions.
The Feynman rules that we need to do this calculation were derived
a long time ago by Feynman
\ref\Fey{R.~P.~Feynman, ``Quantum Theory of Gravitation'', 
{\it Acta Phys. Polon.} {\bf 24} (1963) 697.} and De Witt
\ref\bd{B.~S.~De Witt, ''Quantum Theory of Gravity'', \physrev {162}
{1966} {162}.} (see also
\ref\sannan{S.~Sannan, ''Gravity as the Limit of the Type-II Superstring
Theory'', \physrev {34} {1986} {1749}.}). The vertices involving
gravitons are rather complicated so that we will address the problem 
using a simpler model that captures the important properties
of the calculation. First, we will be assuming that one of the 
gravitons involved in the process is heavy so that it can be treated
as the source for the gravitational field in which the other graviton is 
scattered. Therefore, we will be considering a problem of 
potential scattering.
Our second assumption is that we can treat the `probe graviton'
as a scalar particle. The vertices describing the
coupling of a scalar to gravity are simpler than vertices involving 
gravitons (see {\bd}). 

We consider the 
eleven-dimensional\foot{We are using the signature 
$(-, +, \dots, +)$.}
Einstein action coupled to a massless scalar field
\eqn\ai{
S=\int d^{11} x \sqrt{-g} \left( {1\o  {\k}_{11}^2} R +{1\o 2} g_{\m\n} 
\partial^{\m} \varphi \partial^{\n} \varphi \right), 
}
and we expand the metric around a flat Minkowski background 
\eqn\aii{
g_{\m\n} = \eta_{\m\n} + h_{\m\n}.
}
Here $h_{\m \n}$ is a small fluctuation that 
describes the source graviton field. The action for the 
`probe graviton' $\varphi$
then becomes
\eqn\aiii{
S_{\varphi}= \int d^{11} x
\left( -{1 \o 2}\varphi \bbox \varphi + {1 \o 2} h_{\m\n} 
\left[ \partial^{\m} \varphi \partial ^{\n} \varphi 
-{1\o 2} \eta^{\m\n} \partial_{\l} \varphi \partial^{\l}
 \varphi\right]+ ``hh\varphi \varphi'' \right), }
where we have indicated explicitly the cubic
vertex and only symbolically the quartic vertex whose explicit
form can be found in {\bd}.
The source graviton has vanishing transverse velocity
and it moves with the speed of light in the eleven direction {\bbpt}.
It is described by the Aichelburg-Sexl metric
\ref\as{P.~C.~Aichelburg and R.~U.~Sexl, ''On the Gravitational
Field of a Massless Particle'', {\it Gen. Rel.} {\bf 2}, No.4 (1971) 303.}
whose only non-vanishing component is
$h_{--}$
\eqn\aiv{
h_{--}={15 N_s \o 2 R^2 M_{pl}^9} {1\o r^7} ,
}
where $p^s_-=N_s/R$ is the momentum of the source graviton
and $M_{pl}$ is the Planck mass that is related to the gravitational 
constant by ${\k}_{11}^2 = 16 \pi^5 /M_{pl}^9$.
The action takes then the form
\eqn\aiiixx{
S_{\varphi}= {1 \o 2} \int d^{11} x
\left( -\varphi \bbox \varphi + h^{++} {\partial}_+ 
\varphi {\partial}_+ \varphi+\dots \right). }
In the following we will consider the single graviton exchange 
process in the potential scattering description.
The light-like compactification 
which is described in terms of a non-relativistic Schr\"odinger 
equation is relevant
for finite $N$ M(atrix) theory
\ref\lsa{L.~Susskind, ``Another Conjecture about M(atrix) Theory'',
Stanford preprint SU-ITP-97-11, hep-th/9704080.}.
The space-like compactification is described in terms of a
relativistic Klein-Gordon equation which leads to additional 
terms in the effective potential that are of no relevance for finite $N$.
We will further generalize these results to the double graviton 
exchange describing the relativistic corrections.

\subsec{One Graviton Exchange in the Light-like Compactification} 

In the light-like compactification 
the equation of motion for $\varphi$ that follows from
{\aiiixx} is
\eqn\av{
\left( {1\o 2 m} {\partial^2 \o \partial {\vec x}^2} +i {\partial \o \partial 
\tau} \right) \varphi = V \varphi \qquad {\rm with} \qquad 
V=-{1\o 2m} h_{--}{{\partial_{\tau}}^2}. 
}
Here $m=N_p/R$, $\tau=x^+/2$ and 
we have used that $\varphi$ is in a state of definite $p_-$.
Equation {\av} has the form of a non-relativistic
Schr\"odinger equation for a particle with mass 
$m$ in an external potential $V$. 

The S-matrix can be obtained using the standard quantum
mechanics techniques.
The free propagator for this massive particle is
\eqn\avi{
K_0({\vec x_f},\tau_f; {\vec x_i},\tau_i) =
\theta(\t_f-\t_i) \int {d^9 {\vec p} \o (2 \pi)^9} 
e^{i {\vec p} ( {\vec x_f} -{\vec x_i}) -iE (\tau_f-\tau_i)  },
}
where $(\vec x$, $\vec p$) denotes the nine-dimensional 
transverse space and 
$E={\vec p}^2/2m$ is the light-cone energy. 
The plane wave solutions to the equation of motion 
have a non-relativistic normalization 
\eqn\avii{
\varphi({\vec x},\tau)={1 \o (2 \pi)^{9/2}} e^{{i}({\vec  p} 
{\vec x} -E \tau)}.
}

To first order the S-matrix is given by
\eqn\aviii{
S^{(1)} =-i \int \varphi^*_{{\rm out}} ({\vec x_f},\tau_f)
K_0( {\vec x_f},\tau_f;{\vec x},\tau) 
V({\vec x},\tau) K_0({\vec x},\tau; {\vec x_i},\tau_i) 
\varphi_{\rm in} ({\vec x_i},\tau_i) d{\vec x_f} d{\vec x} 
d{\vec x_i} d \tau
.}
After evaluating the different integrals we get\foot{Here and in the 
following we omit the factor $2 \pi i \delta(E_f-E_i)$ that 
expresses energy conservation from all the scattering amplitudes. }
\eqn\aix{
S^{(1)}=- {N_s N_p \o R^3 M_{pl}^9} {v^4 \o 8 (2\pi)^5}
{1 \o | {\vec k}|^2} ,
}
where ${\vec k}={\vec p_f}-{\vec p_i}$ 
is the nine-dimensional momentum transfer.
The effective potential is obtained by Fourier-tranforming
\eqn\ax{
{V_{eff}^{(1)}} =-{15 \o 16} {N_s N_p \o R^3 M_{pl}^9} {v^4 \o r^7}. 
}
This is the result that was computed in 
{\bfss}, 
\ref\beco{D.~Berenstein and R.~Corrado, ``M(atrix) Theory in
Various Dimensions'', \pl {406} {1997} {37}, hep-th/9702108.}
and {\bbpt}.

\subsec{One Graviton Exchange in the Space-like Compactification}
The previous calculation can be repeated for a 
space-like compactification. 
In this case the equation of motion for $\varphi$ has the form
of a relativistic Klein-Gordon equation:
\eqn\bi{
\left( {\partial^2 \o \partial {\vec x}^2}-
{\partial^2 \o \partial x_0^2}+ {m^2}    \right) 
\varphi = V \varphi
\qquad {\rm with} \qquad 
V=-h_{--}({\partial}_{x_0}-m)^2,}
where $m$ is the eleven-component of the momentum.
The solutions to the free equation of motion are now plane-waves with
a relativistic normalization
\eqn\bii{
\varphi(x)={1\o (2\pi)^{9/2}}{1\o \sqrt{2E}} e^{ipx} ,
}
where 
$E^2={\vec p}^2+m^2$.

The S-matrix to first order is now
\eqn\aviixx{
S^{(1)} =-i \int \varphi^*_{{\rm out}} (y)
V(y) 
\varphi_{\rm in} (y) d^{10}y.
}
After performing the integration and a
Fourier transformation we obtain
\eqn\biv{
V_{eff}^{(1)} =-{N_p \o R } {h_{--} \o 2} {(\sqrt{1-v^2} -1)^2 \o 
\sqrt{1-v^2} }. }
This result agrees with {\bbpt}.
While in the light-like compactification the S-matrix contained only a 
$v^4$-term {\ax}, here we obtain a series in terms of $v$.
The difference is due to the fact that in the second calculation 
$\varphi$ obeys relativistic kinematics, while in the light-like
compactification $\varphi$ obeys a non-relativistic Sch\"odinger equation.  
These additional terms have no relevance for finite $N$ M(atrix) theory
and are subleading in the large $N$ expansion.

\newsec{General Relativity Corrections}
\subsec{Double Graviton Exchange in the Light-like Compactification}
Corrections coming from general relativity are described by the
diagonal term in the table appearing in {\bbpt}. They have 
integer powers in ${\k}_{11}$.
The simplest example of a correction of this type is the 
$v^6/r^{14}$-term appearing in M(atrix) theory at two-loops.
It can be obtained from the double graviton exchange or equivalently
a second order scattering process.
Here we will evaluate the contribution involving 
two cubic vertices of the type {\av}.
To second order the S-matrix described by 
{\av } is given by
\eqn\axi{
\eqalign{
S^{(2)} =-\int & {\varphi}_{\rm out}^* ({\vec x_f},\tau_f)
K_0({\vec x_f}, \tau_f; {\vec x'}, \tau') 
V({\vec x'},\tau') K_0( {\vec x'},\tau'; {\vec x},\tau ) \cr 
&  V({\vec x}, \tau) K_0({\vec x},\tau; {\vec x_i}, \tau_i) 
\varphi_{\rm in} ({\vec x_i},\tau_i) d{\vec x_i} 
d{\vec x} d \tau d{\vec x'} d \tau' d {\vec x_f}.  \cr
}}
We can evaluate the ${\vec x_i}$, ${\vec x_f}$, ${\vec x}$ 
and ${\vec x'}$ integrals exactly. From the 
$\tau$, $\tau '$ integrals we get the contribution
\eqn\axii{
\int_{-\infty}^{+\infty} d \tau e^{ i \tau (E_f-E_i) }
\int_{0}^ {+\infty} dt e^{ i t (E_f -{{\vec p}^2 \o 2m} )},
}
where we have transformed to a new integration variable $t=\tau-\tau'$.
The first integral again gives energy conservation but the second 
integral does not converge. We can evaluate it using a trick of 
Feynman and Hibbs 
\ref\fh{R.~P.~Feynman and A.~R.~Hibbs, ''Quantum Mechanics 
and Path Integrals''
McGraw-Hill Book Company, 1965.}. 
Expression {\axii} can be written as
\eqn\axiii{
2 \pi \delta(E_f-E_i) \lim_{\e \to 0} {i \o E_f-E+ i \e}, }
where $E={\vec p}^2/{2m}$.
The S-matrix to second order takes then the form
\eqn\axiv{
S^{(2)} = {N_p^3 N_s^2 \o R^7 M_{pl}^{18}} {v^8 \o 32}
\lim_{\e \to 0} \int {d^9 {\vec p} \o (2 \pi )^{10}} 
{1 \o | {\vec p_f}-{\vec p} |^2} 
{1\o | {\vec p_i}-{\vec p} |^2 }{1\o {\vec p}^2 -{\vec p_f}^2 -i \e}}
This is a standard integral that appears in the computation
of radiative corrections to Coulomb scattering
\ref\akba{A.~I.~Akhiezer and V.~B.~Berestetskii, 
``Quantum Electrodynamic'', Interscience Monographs and Texts
in Physics and Astronomy'', Vol. XI, New York, 1965.} and we solve it
explicitly in the appendix. The leading non-analytic 
behavior of this integral is
\eqn\axixi{
I={{\pi}^6 \o 256}{|\vec k|^{5}\o p_f^2}.}
In order to Fourier transform this expression we need: 
\eqn\zxvi{
\int {d^9 {\vec k}} e^{i {\vec k} {\vec r} }
|{\vec k}|^{5} = -2^{15} 3^3 5^2 {{\pi}^4 \o r^{14}} .} 

Up to a constant the 
final result for the effective potential is
\eqn\axxi{
{V_{eff}^{(2)}} \propto -{225 \o 64} {N_s^2 N_p \o R^5 M_{pl}^{18}} 
{v^6 \o r^{14}}. 
}
This shows that this diagram indeed gives a contribution to the 
$v^6/r^{14}$-term. 
To get the exact numerical coefficient one would have to 
evaluate all possible double graviton exchange diagrams.
Moreover 
from the 
appendix we observe that the imaginary part of this amplitude
is infrared divergent
\foot{Infrared divergences in quantum gravity can be 
treated in the same manner as in quantum electrodynamics
\ref\wein{S.~Weinberg, ``Infrared Photons and Gravitons'', 
{\it Phys. Rev.} {\bf 140} (1965) 516.}.}. 
This is familiar from the calculation
of radiative corrections to Coulomb scattering in four-dimensions
{\akba}. To cancel this divergence one gives a mass to the photon and
the sum over all polarizations gives a finite amplitude.
Polarizations for the probe graviton will have to be taken
into account here as well at some point.
This calculation seems much harder because graviton vertices are
rather complicated. The easiest way to proceed might be to compute this
four-graviton amplitude using  type IIA string theory as 
we do in the next section. 
Of course, we expect the loop amplitude to reproduce the 
result of the classical
calculation of {\bbpt}. 
This is similar to the situation appearing in the
four-dimensional Newton 
potential\foot{We thank Steven Weinberg for pointing out this reference 
to us.}
\ref\Dono{J.~F.~Donoghue, ``Leading Quantum Correction 
to the Newton Potential'', {\it Phys. Rev. Lett.} {\bf 72} (1994) 2996.}. 
Here general relativity corrections can be computed 
through graviton scattering amplitudes or by a classical calculation in
which the zero component of the Schwarzschild metric is expanded.

\subsec{The Covariant Four-Graviton Amplitude}
It is curious to see that in the space-like compactification
the general relativity correction of the previous section can be 
computed from the
covariant four-graviton amplitude 
{\gsb} {\tsru}.
The sum of the tree level and one-loop contributions to the on-shell 
four-graviton amplitude in $D$ dimensions 
is given by {\gsb}: 
\eqn\zi{
{\rm (kinematical \;\; factor)}\times
\left( {1\o stu} 
{\Gamma(1-{\a' \o 2} s) \Gamma( 1-{\a' \o 2} t) \Gamma(1- {\a' \o 2} u) 
\o \Gamma(1+{\a' \o 2} s) \Gamma(1+{\a' \o 2} t) \Gamma(1+{\a' \o 2} u) } 
+c_1 g^{(1)} +\dots \right) , 
}
where $c_1$ is a calculable number determined by unitarity 
\ref\sata{N.~Sakai and Y.~Tanii, ``One-Loop Amplitudes and Effective
Action in Superstring Theories'', \np {287} {1987} {457}.}
and $s$, $t$ and
$u$ are the Mandelstam variables 
\eqn\zx{
\eqalign{
s & =-(p_1+p_2)^2 , \cr 
t & = -(p_1+p_4)^2 ,\cr 
u & =- (p_1+p_3)^2 .\cr }
}
Here
$1$ and $2$ are incoming particles while $4$ and $3$ are outgoing and
$p_i$ describe their corresponding momenta.
The first term in formula {\zi} represents the tree 
level contribution while the 
one-loop part of the amplitude $g^{(1)}$ is given by: 
\eqn\zii{
g^{(1)}= {\k_{10}^2 \o \a'}  \int {d^2 \t \o ({\rm Im} \t)^2 }  
F(\t) \left[ F_2 (a,\t)\right] ^{10-D} , 
}
where $a=\sqrt{\a'} / R$ is a dimensionless parameter. 
The expression for $F(\t)$ can be found 
in {\gsb}. 
We are interested in the low-energy limit of this amplitude which
corresponds to the zero slope limit $\a' \to 0$.
The one-loop contribution is divergent for 
$D \geq 8$ (here we consider $D=11$).
The finite part of {\zi} reduces to  
\eqn\ziii{
{\rm (kinematical \; \; factor ) }\times 
\left( {1\o stu} + c_1 g^{(1)}_0 +\dots \right) , 
}
where 
\eqn\ziv{
g_0^{(1)}= \k_D^2 c (\g) \left[ 
I_{\g} ( s,t) +I_{\g} (t,s) + I_{\g} ( s,u) +I_{\g} (u,s) 
+I_{\g} (t,u) +I_{\g} (u,t) \right], 
}
is the asymptotic value of $g^{(1)}$ and $\g=D/2-4$. 
Here $c(\g)$ is a constant 
determined by the dimensionality of the space-time 
\eqn\zv{
c(\g) ={ 1\o 4} \left( { \pi \o 4} \right)^{\g+1/2} 
\left[ \sin( \pi \g) \Gamma(\g +5/2) \right] ^{-1}. 
}
The function $I(\g)$ is 
\eqn\zvi{
I_{\g} (s,t) = t^{\g+1} \int_0^1 {(1-x)^{\g+1} dx \o sx-t(1-x) } . }

In $D=11$ we get $\g=3/2$ so that $c(\g)$ is: 
\eqn\zvii{c(3/2)= -{\pi^2 \o 384}, 
}
and $g_0^{(1)}$ 
\eqn\zviii{
g_0^{(1)} = -{\pi^2 \o 384} \k_{11}^2 
\left[ I_{3/2} ( s,t) +I_{3/2} (t,s) + I_{3/2} ( s,u) +I_{3/2} (u,s) 
+I_{3/2} (t,u) +I_{3/2} (u,t) \right]. }
The integrals $I_{3/2}$ can all be evaluated exactly but 
give rather complicated functions. Expanding around $t=0$ 
we are interested in the leading non-analytic\foot{The analytic 
terms correspond to
contact terms after Fourier transforming.} term in $t$.  
It is given by: 
\eqn\zix{
g_0^{(1)} = 
i \k_{11}^2  {\pi^3 \o 384} {t^{5/2} \o s} +{\cal O} (t^{7/2})  . } 

The kinematical factor in {\zi} can be computed from the 
tree level contribution of this formula.
We consider the case where the first graviton has vanishing 
transverse velocity, i.e. $p_1=(M,0,M)$, where 
$M=N_s/R$ is the mass. The second incoming graviton has 
$p_2=(E_2 , {\vec p_2},m)$,
where $E_2^2= {\vec p_2}^2+m^2$ is the relativistic 
energy, ${\vec p_2}$ is the 
nine-dimensional transverse momentum and $m=N_p/R$ is the mass.
To leading order in the velocity we have 
\eqn\zxi{ \eqalign{
s & = -M m {\vec v^2} +\dots, \cr 
t &  = -| {\vec k} |^2 +\dots,  \cr 
u &  = -s-t. \cr }}
Here we have set $|{\vec p_2}|=mv$ to leading order in $v$. 
As can be easily checked 
the result for the tree level contribution
computed in {\bfss} and {\beco} is reproduced with
a kinematical factor $-\k_{11}^2 s^2 u^2$. 
The tree level
amplitude then becomes 
\eqn\zxii{
A_{\rm tree}= \k_{11}^2 { s^2 \o t}=-\k_{11}^2 M^2 m^2 
{v^4 \o |{\vec k}| ^2} . }
In order to obtain the potential we are interested in the 
Fourier transformed of this expression with respect to 
the momentum transfer. Using the relation
\eqn\zxiii{
 \int { d^9 {\vec k} \o (2 \pi)^9} 
{e^{i{\vec k} {\vec r} } \o |{\vec k}|^2 } 
= {15 \o 2 (2 \pi)^4 } {1 \o r^7},  
}
we obtain  
\eqn\zxiv{V_0(r) ={1\o 2 \pi R} \int 
{d^9 {\vec k} \o (2 \pi)^9} e^{i {\vec k} {\vec r} } 
A_{\rm tree} = {15 \o 16} {N_p N_s \o R^3 M_{pl}^9} {v^4 \o r^7}. 
}
In order to obtain this result 
we have included a normalization factor of $1/\sqrt{2E}$ 
for each state. 

Using the above kinematical factor we can 
obtain the form of the one-loop contribution to the potential. 
The one-loop amplitude is given by: 
\eqn\zxv{
A_{\rm 1-loop}= -\k_{11}^4 c_1 {\pi^3 \o 384} M^3 m^3 v^6 
| {\vec k}|^{5}. }
We Fourier transform this expression using {\zxvi}. 
Up to a constant the result for the one-loop 
contribution to the potential is then: 
\eqn\zxvii{
V_1(r) \propto N_p {225 \o 64} {N_p^2 N_s \o R^5 M_{pl}^{18} } 
{v^6\o r^{14}}. 
}
We observe that the covariant amplitude differs by a factor 
$N_p$ from the result calculated in
{\bbpt} using a classical picture. The difference is due to the fact 
that the covariant amplitude integrates over all eleven-dimensional
momenta running in the loop while in the M(atrix) theory calculation 
only states with fixed $p_-$ are allowed to run through the loop.
Both amplitudes are related\foot{We thank David Gross 
and Lenny Susskind for 
discussions on this.} 
by a phase factor times
$N_p$. This has to be taken into account when loop
diagrams are calculated while tree diagrams naturally agree.

\newsec{Quantum Gravity Corrections}
M(atrix) theory not only allows us to compute corrections
coming from general relativity but the much harder 
quantum gravity corrections.
They have, in general, fractional 
powers of $\kappa$ because a dimensionful parameter, the cutoff, 
is present in the calculation. 
An example of a correction of this sort is the $v^8/r^{18}$-term
appearing in M(atrix) theory at two-loops.
The result for this term in the effective potential can
be easily obtained by a simple extension of the 
calculation performed in
\ref\bebe{K.~Becker and M.~Becker, ``A Two-Loop Test of M(atrix) 
Theory'', \np {506} {1997} {48}, hep-th/9705091.}
\eqn\cvii{
S_{v^8} = {3\cdot 5^2 \cdot 7^2 \cdot 59  \o 2^{11}} 
{N_s^2 N_p \o R^7 M^{24}} {v^8 \o r^{18}}. 
}

Susskind conjectured that this term is related to a 
four graviton amplitude involving the $R^4$ vertex
of M-theory {\leta}
\foot{Some discussion on this appeared in
\ref\beg{P.~Berglund and D.~Minic, ``A Note on Effective Lagrangians
in Matrix Theory'',  preprint EFI-97-37, hep-th/9708063.}
\ref\ser{M.~Serone, ``A Comment on the $R^4$  
Coupling in M(atrix) Theory'',
preprint UVA-WINS-WISK-97-13, hep-th/9711031.}
\ref\vakr{E.~Keski-Vakkuri and P.~Kraus, ``Short Distance Contributions
to Graviton-Graviton Scattering: Matrix Theory versus Supergravity'', 
Caltech preprint CALT-68-2148, hep-th/9712013.}
\ref\bgl{V.~Balasubramanian, R.~Gopakumar and F.~Larsen, ``Gauge
Theory, Geometry and the Large N Limit'', Preprint HUTP-97/A095, 
hep-th/9712077.}.}
From the M-theory point of view the following is known.
The leading term in the M-theory effective action is the conventional 
Einstein term. However there could be corrections coming from 
higher dimensional operators: 
\eqn\ci{R^2+R^3+R^4 +\dots.}
Fradkin and Tseytlin 
\ref\ft{E.~S.~Fradkin and A.~A.~Tseytlin, 
``Quantum Properties of Higher Dimensional and Dimensionally Reduced
Supersymmetric Theories'', \np {227} {1983} {252}.} 
have shown that 
\eqn\cii{R^m=0 \qquad {\rm for } \qquad m=0,1,2,3 }
at one loop. The one-loop correction to the effective action takes then the 
form:
\eqn\ciii{S \propto \int d^{11} x \sqrt{ -g} \L_{11}^3 R^4 ,}
where $\L_{11}$ is a dimensionful cutoff. 
This cutoff can be determined from the type IIA string theory, 
as was done in 
\ref\gv{M.~B.~Green and P.~Vanhove, ``D-Instantons, Strings and 
M-Theory'', \pl {408} {1997} {122}.} and
\ref\ggv{M.~B.~Green, M.~Gutperle and P.~Vanhove, 
``One Loop in Eleven Dimensions'', \pl {409} {1997} {177}.}. 
These authors
decompactified the four-graviton 
amplitude of $d=9$ closed superstring theory to calculate the 
$R^4$ interaction in eleven dimensions: 
\eqn\civ{
S={1\o 2 {\k}_{11}^2} \int \sqrt{-g} R 
+{ c \o {\k}_{11}^{2/3} }\int 
\sqrt{-g} t_8 t_8 R^4,} 
where $c$ is a number that can be extracted from {\gv}.
It is plausible to use this result to
compute the analog of {\cvii} from the supergravity point of view. 
While it seems hard to do a precise calculation because we 
expect polarizations and fermions to be relevant, it is
not too hard to see that the four graviton amplitude involving 
the vertex {\civ} indeed gives a contribution to {\cvii}.
We can use the probe-source picture of the previous 
sections\foot{We thank Joe Polchinski for discussions on this.}.
Higher-order local curvature invariants in eleven-dimensional 
supergravity do not change the form of $h_{--}$ 
\ref\host{G.~T.~Horowitz and A.~R.~Steif, 
``Space-Time Singularities in String Theory''
{\it Phys. Rev. Lett.} {\bf 64} (1990) 260.}. 
The only effect of such a term is to change the 
action of the probe, so that the equation of motion for $\varphi$ 
involves a new potential: 
\eqn\cv{V({\vec x}, \tau) \varphi \propto
\left( {\partial^2 h_{--} \o \partial {\vec x}^2 } \right)^2 
\partial_{\tau}^4 \varphi+\dots }
This potential comes from the term $(R_{-r-r})^2 (R_{+i+j})^2 $ 
of the interaction vertex. The dots indicate other contributions 
in which we are not interested.  Up to a constant the effective
action from the supergravity calculation is: 
\eqn\cvi{
S_{v^8} \propto {N_p^3 N_s^2 \o R^7 M_{pl}^{24}} {v^8\o r^{18}} .}

The dependences on $v$, $r$ $N_s$, $R$ and $M_{pl}$ agree 
with {\cvi}, but we seem to find a disagreement for the 
dependence on $N_p$. 
While the M(atrix) theory result is linear in $N_p$, the M-theory
result is proportional to $N_p^3$. It seems plausible that the origin
of this discrepancy is again the fact that in the
M-theory calculation states with eleven-dimensional
momentum are running through the loops, while in the M(atrix) theory
calculation these states have a fixed $p_-$. This changes the $N_p$
dependence of the result as we have seen in the previous sections.  
A more careful analysis including polarizations and fermions would have
to be done to check if the numerical coefficient agrees. 
While the M(atrix) theory calculation was an easy extension
of our previous calculation {\bebe}, the DLCQ M-theory
counterpart seems to be much harder
\ref\hepo{S.~Hellerman and J.~Polchinski, ``Compactification 
in the Lightlike Limit'', preprint NSF-ITP-97-139, hep-th/9711037.}.

One may want to ask at this moment: why should the supergravity 
calculation agree with the M(atrix) theory calculation at all, 
especially because of the discrepancies found in the recent papers
\ref\dira{M.~Dine and A.~Rajaraman, ``Multigraviton Scattering in the 
Matrix Model'', preprint SLAC-PUB-7676, hep-th/9710174.} 
\ref\doog{M.~R.~Douglas and H.~Ooguri, ``Why Matrix Theory is Hard'', 
preprint LBL-40889, hep-th/9710178.} or why does the relativistic
correction come out correctly in the calculation performed in {\bbpt}?
Probably the most natural explanation\foot{We thank Lenny Susskind for
discussions on this.}
would be that the quantities 
computed in {\dira} and {\doog} receive corrections in the large $N$ limit,
while the quantity computed in {\bbpt} might obey a non-renormalization
theorem that guarantees that the supergravity result is reproduced 
by M(atrix) theory for finite $N$.
Our goal for the near future should then be to find out which quantities
are not renormalized in the large $N$ limit so that they 
can be computed with finite $N$ M(atrix) theory.
Maybe an argument along the lines of 
\ref\malda{J.~Maldacena, ``The Large N Limit of Superconformal 
Field Theories and Supergravity'', preprint HUTP-98-A097, 
hep-th/9711200.} can be carried out here as well.

\vskip 1cm
\noindent {\bf Note Added}

\noindent 
These results have been presented in several seminars in October 
and November of this year and at the 3rd Workshop on Recent 
Developments
in Theoretical Physics at CERN on December 9. 
  
\vskip 1cm
\noindent {\bf Acknowledgements}

\noindent
We would like to thank M.~Dine, W.~Fischler, M.~Green,   
D.~Gross, J.~Maldacena, J.~Polchinski, A.~Rajaraman, 
J.~Schwarz, S.~Shenker and L.~Susskind for useful discussions.
This work was supported by the U.S. Department of Energy 
under grant DE-FG03-92-ER40701.

\newsec{Appendix}
In this appendix we are going compute explicitly the integral: 
\eqn\appi{
I=\lim_{\epsilon\to 0} \int 
d^9{\vec p} {1 \o \mid {\vec p}_f -\vec p \mid^2} 
{ 1\o \mid {\vec p}_i - {\vec p} \mid^2} 
{1 \o {p}^2 - {p}_f^2 - i \e}, }
where $p=|{\vec p}|$ and the same for ${\vec p}_f$. The above 
integrand can have poles when the denominator vanishes. 
In order to regularize them we introduce a "graviton mass" $\sigma$.  
At the end of the calculation we will be interested in the 
leading order behavior in terms of $\sigma$. 
Therefore we would like to compute the integral: 
\eqn\appii{
I=\lim_{\epsilon\to 0} \int 
d^9{\vec p} {1 \o \mid {\vec p}_f -\vec p \mid^2 + \sigma^2} 
{ 1\o \mid {\vec p}_i - {\vec p} \mid^2+\sigma^2 } 
{1 \o  {p^2} - {p}_f^2 - i \e}}
to leading order in $\s$. First it is convenient to use the formula: 
\eqn\appiii{
{ 1\o ab} ={1 \o 2} \int_{-1}^1 
{ dz \o \left( a(1+z)/2+b(1-z)/2 \right)^2 }, 
}
to write: 
\eqn\appiv{
{ 1\o \mid {\vec p} -{\vec p}_f \mid^2 \mid {\vec p} -{\vec p}_i \mid^2} 
=-{1 \o 4 } \int_{-1}^1 
 {1 \o \L}  {\partial \o \partial \L}\left[
 { 1\o ( {\vec p} -{\vec P})^2 +\L^2 }\right]
d z , 
}
where 
\eqn\appv{
\eqalign{
& {\vec P}  = {1\o 2} \left[  (1+z) {\vec p}_f +(1-z) 
{\vec p}_i \right],\cr
& \L^2  = \sigma^2 +  { 1\o 4} (1-z^2) k^2 , \cr   }}
and $\vec q=\vec p_f - \vec p_i$ is the momentum transfer. 
Using this identity $I$ can be written in the form: 
\eqn\appvi{
I=-{1 \o 4} \int_{-1}^1 dz {1\o \L} {\partial \o \partial \L} {\tilde I}
\qquad {\rm with} \qquad {\tilde I} = \lim_{\e \to 0} \int d^9{\vec p} 
{1\o ({\vec p} -{\vec P})^2 +\L^2} {1 \o p^2-p_f^2 - i \e}. }
We will compute the real and the imaginary 
part of the above integral separately. In order to do this 
we use the identity: 
\eqn\appvii{
{1\o  p^2 -p_f^2 - i \e} = 
P.P. \left( {1\o  p^2 -p_f^2 }\right) +
i\pi \d(   p^2 -p_f^2) , 
}
where $P.P.$ denotes the principal part and $\d$ is the Dirac 
delta function. we will first begin with the imaginary part. 
It is given by: 
\eqn\appviii{
{\rm Im } {\tilde I} = \pi 
\int 
d^9{\vec p} {\d(    p^2 -p_f^2)
\o ({\vec p} -{\vec P})^2 +\L^2}=-{ \pi^4  \o 12} {p_f^6 \o P }
\log \left( { \L^2 +( {\vec P} +{\vec p}_f)^2 \o \L^2+( {\vec P}-
{\vec p}_f)^2 } \right) .  }
Inserting in {\appvi} we can get the result for the imaginary 
part of $I$ to leading order in $\sigma$: 
\eqn\appix{
{\rm Im} I = {\pi^5 \o 6} {p_f^5 \o q^2} \log \left( {\sigma^2 \o q^2}
\right)+ {\cal O} ( \sigma^2).}

The real part of $\tilde I$ is  
\eqn\appx{
{\rm Re} {\tilde I} = \int d^9 {\vec p} 
{ 1\o ({\vec p} - {\vec P})^2+\L^2} P.P. \left( 1\o  p^2 -p_f^2
\right).
}
to get 
\eqn\appxii{{\rm Re} {\tilde I} =\int_0^{\infty} dk  \int d^9 {\vec p} 
{ \sin k(p^2 - p_f^2) \o ({\vec p} - {\vec P})^2+\L^2} }
This integral is straightforward to compute just a bit lengthy. 
Here we are only going to write down the terms 
of ${\rm Re} I$ in which we are interested in. These are the terms that 
in the region for small momentum transfer are non-analytic. To 
leading order in $\sigma$ they are: 
\eqn\appxiii{
{\rm Re} I = {\pi^6 \o 4}{ p_f^4 \o  ( 4 p_f^2-q^2)^3}
 |q|^{5}\left( 1-{q^2 \o 6 p_f^2}+
{q^4 \o 80 p_f^4}  \right) +\dots}

The leading non-analytic behavior of $I$ in terms of ${\vec q}$ 
is then given by: 
\eqn\appxiv{
I={\pi^6 \o 256 } {|q|^{5} \o p_f^2} + i 
{ \pi^5 \o 6} {p_f^5 \o q^2} \log \left( {\sigma^2 \o q^2} \right)+\dots}

\listrefs

\end